\documentstyle[aps,prl,twocolumn,psfig,floats]{revtex}
\begin{document}
\preprint{\vbox{\hbox{UCB-PTH-00/03}},
  \vbox{LBNL-45023}}
\draft
\wideabs{
\title{The Dark Side of the Solar Neutrino Parameter
Space\cite{acknowledgements}}
\author{Andr\'e de Gouv\^ea}
\address{CERN - Theory Division, CH-1211 Geneva 23, Switzerland}
\author{Alexander Friedland and Hitoshi Murayama}
\address{
Department of Physics,
University of California,
Berkeley, CA~~94720, USA\\
Theory Group,
Lawrence Berkeley National Laboratory,
Berkeley, CA~~94720, USA}

\date{\today}
\maketitle

\begin{abstract}
Results of neutrino oscillation experiments have always been presented
on the $(\sin^{2} 2\theta, \Delta m^{2})$ parameter space for the case
of two-flavor oscillations.  We point out, however, that this
parameterization misses the half of the parameter space $\frac{\pi}{4}
< \theta \leq \frac{\pi}{2}$ (``the dark side''), which is physically
inequivalent to the region $0 \leq \theta \leq \frac{\pi}{4}$
(``the light side'') in the presence of matter effects.  The MSW solutions
to the solar neutrino problem can extend to the
dark side, especially if we take the conservative attitude to allow
higher confidence levels, ignore some of the experimental results in the
fits, or relax theoretical predictions.  Furthermore, even the so-called
``vacuum oscillation'' solution distinguishes the dark and the
light sides.  We urge experimental
collaborations to present their results on the entire parameter space.
\end{abstract}
}
\vskip 0.3in

In the Standard Model of particle physics, neutrinos are strictly
massless.
Recently, however, the Super-Kamiokande collaboration studied atmospheric
neutrinos and reported a strong evidence for neutrino oscillations
\cite{SKatm}, and hence a finite neutrino mass.  The most likely
interpretation of their data is the oscillation between $\nu_{\mu}$
and $\nu_{\tau}$.  This made it also natural to interpret another
long-standing issue in neutrino physics, the deficit of the solar
$\nu_{e}$ flux \cite{solarreview}, in terms of neutrino oscillations.
However, the solar neutrino deficit has not been regarded as
convincing evidence for neutrino oscillations in the community.  The
reason is probably multifold but two main objections are the
following. (1) Neutrino experiments are so difficult that it is
possible that some of the data are not entirely correct.  (2) The
physics of the Sun is so complex that the neutrino flux calculations
in the Standard Solar Model (SSM) may have underestimated the theoretical
uncertainties.

To resolve this situation, a new generation of solar
neutrino experiments, such as Super-Kamiokande, SNO, Borexino, GNO,
KamLAND, etc, is looking for an evidence for solar neutrino
oscillations without relying on the SSM in
well-understood experimental environments.  They aim not only at
establishing oscillations but also at overdetermining the
solution in the next few years.  Such data will eventually supersede
data from the past experiments.  It is therefore important to analyze
the future data without too much prejudice based on the past data.

In this letter, we point out that the study of neutrino oscillations on
the $(\Delta m^{2}, \sin^{2} 2\theta)$ parameter space done
traditionally is incomplete, since it covers only the range $0 \leq
\theta \leq \frac{\pi}{4}$ (``the light side'').  Indeed, some of the solutions to
the solar neutrino puzzle extend to the other half of the parameter
space $\frac{\pi}{4} < \theta \leq \frac{\pi}{2}$, which we call ``the dark side,'' and
hence it is phenomenologically necessary to include both halves of the
parameter space.  This is especially true once one employs a more
conservative attitude which either allows higher confidence levels,
ignores some of the experimental data (especially Homestake \cite{Cl}),
or relaxes the theoretical prediction on the $^{8}$B flux.  

Neutrino oscillations occur if neutrino mass eigenstates are different
from neutrino weak eigenstates.  Assuming that only two neutrino
states mix, the relation between mass eigenstates ($\nu_{1}$ and
$\nu_{2}$) and flavor eigenstates (for example $\nu_{e}$ and
$\nu_{\mu}$) is simply given by
\begin{eqnarray}
\label{masseigen}
|\nu_1\rangle=\cos\theta|\nu_e\rangle-\sin\theta|\nu_{\mu}\rangle,
\nonumber \\
|\nu_2\rangle=\sin\theta|\nu_e\rangle+\cos\theta|\nu_{\mu}\rangle,
\end{eqnarray}
where $\theta$ is the vacuum mixing angle.  The mass-squared
difference is defined as $\Delta m^2\equiv m_2^2-m_1^2$.  We are
interested in the range of parameters that encompasses all physically
different situations.  First, observe that Eq.~(\ref{masseigen}) is
invariant under $\theta \rightarrow \theta+\pi$, $|\nu_e\rangle
\rightarrow -|\nu_e\rangle$, $|\nu_\mu\rangle \rightarrow
-|\nu_\mu\rangle$, and hence the ranges $[-\frac{\pi}{2},\frac{\pi}{2}]$ and
$[\frac{\pi}{2},\frac{3\pi}{2}]$ are physically equivalent.  Next, note that it is
also invariant under $\theta \rightarrow -\theta$, $|\nu_\mu\rangle
\rightarrow -|\nu_\mu\rangle$, $|\nu_2\rangle \rightarrow
-|\nu_2\rangle$, hence it is sufficient to only consider $\theta\in
[0,\frac{\pi}{2}]$.  Finally, it can also be made invariant under $\theta
\rightarrow \frac{\pi}{2} - \theta$, $|\nu_\mu\rangle \rightarrow
-|\nu_\mu\rangle$ by relabeling the mass eigenstates $|\nu_1\rangle
\leftrightarrow |\nu_2\rangle$, {\it i.e.} $\Delta m^2 \rightarrow
-\Delta m^2$.  Thus, we can take ($\Delta m^2>0$) without loss of
generality.  All physically different situations are obtained by
allowing $0\leq \theta\leq \frac{\pi}{2}$.

For the case of oscillations in the vacuum, the survival probability
is given by
\begin{equation}
        P(\nu_{e}\rightarrow \nu_{e}) = 1 - \sin^{2} 2\theta
                \sin^{2} \left( 1.27 \frac{\Delta m^{2}}{E}L \right).
        \label{eq:P}
\end{equation}
Here, $\Delta m^{2}$ is given in
eV$^{2}/c^{4}$, $E$ in GeV, and $L$ in km.  In this case the
oscillation phenomenon can be parameterized by $\Delta m^{2}$ and
$\sin^{2} 2\theta$, since $\theta$ and $\frac{\pi}{2}-\theta$ yield
identical physics.  Therefore we can restrict ourselves to $0\leq
\theta\leq \frac{\pi}{4}$, and use the parameter space $(\Delta
m^{2}, \sin^{2} 2 \theta)$ without any ambiguity.  This is indeed an
adequate parameterization for reactor antineutrino oscillation
experiments, short-baseline accelerator neutrino oscillation
experiments, and $\nu_{\mu} \leftrightarrow \nu_{\tau}$ atmospheric
neutrino oscillation experiments.

In certain cases, however,
neutrino-matter interactions can dramatically change the oscillation
probability \cite{MSW}.  These matter effects are particularly
important in explaining the solar $\nu_e$ flux deficit in terms of
neutrino oscillations.
In the presence of matter effects, Eq.~(\ref{eq:P}) is modified to
\begin{eqnarray}
        & & P(\nu_{e}\rightarrow \nu_{e}) =
        P_{1} \cos^{2} \theta + (1-P_{1}) \sin^{2} \theta \nonumber \\
        & &
        - \sqrt{P_{c} (1-P_{c})} \cos 2\theta_M \sin 2 \theta
        \cos \left( 2.54 \frac{\Delta m^{2}}{E}L + \delta \right),
        \label{eq:Pmatter}
\end{eqnarray}
where $P_c$ is the hopping probability, $\theta_M$ is the mixing angle
at the production point, $P_1 = P_c \sin^2 \theta_M + (1-P_c) \cos^2
\theta_M$, 
and $\delta$ is a phase induced
by the matter effects, which is not important for
our purposes.  See
Ref.~\cite{seasonal,us} for notation.  Because
$P_{1}$ depends on 
$\Delta m^{2} \cos 2\theta$, 
the half of the parameter space
traditionally considered $0 \leq \theta \leq \frac{\pi}{4}$ (the light
side) is physically inequivalent to the other half $\frac{\pi}{4} <
\theta \leq \frac{\pi}{2}$ (the dark side).  However, all data analysis have
been reported on the $(\Delta m^{2}, \sin^{2} 2 \theta)$ plane
with positive $\Delta m^{2}$ only 
for solar neutrino experiments and hence only half of the parameter
space has been analyzed.  Even though the dark side has been studied
in the context of three-flavor \cite{FLMP} and four-flavor
\cite{4flavor} neutrino oscillations, the importance of studying both
halves for the simplest case of two-flavor oscillations has been
largely ignored in the literature.  There also appeared to be a
misconception in the literature that physics was discontinuous at
maximal mixing $\theta=\frac{\pi}{4}$.  For instance, matter effects in the
Earth were once thought to disappear as the mixing approached maximal.
However, the authors of Ref.~\cite{Guth_Randall} emphasized that the
matter effects remain important even for the maximal mixing, and the
present authors further showed that physics is completely continuous
beyond $\theta=\frac{\pi}{4}$ \cite{us}.  One can still
retain the dark side with only $0\leq \theta\leq \frac{\pi}{4}$ if
a separate parameter space with $\Delta m^{2} < 0$ is added.  This is
indeed what Super-Kamiokande did in the case of $\nu_\mu
\leftrightarrow \nu_s$ oscillations of atmospheric neutrinos
\cite{atmossterile}.  However, as argued in \cite{us}, it is more
natural to use $0\leq \theta\leq \frac{\pi}{2}$ with the fixed sign of
$\Delta m^{2}$ to exhibit the continuity of the physics between the
two halves of
the parameter space.

Part of the reason why the dark side has been neglected in the
literature is that it is impossible to obtain $\nu_{e}$ survival
probabilities less than one half when the two mass eigenstates are 
incoherent, {\it i.e.}\/, when
the last term in Eq.~(\ref{eq:Pmatter}) is absent.
(This occurs in the so-called
``MSW region'' $10^{-8} \lesssim \Delta m^{2} \lesssim 
10^{-3}~{\rm eV}^{2}$ \cite{decoherence}.) 
Indeed, the data from the Homestake
experiment \cite{Cl} used to be about a quarter of the SSM
prediction, and this could have been an argument for dropping
the dark side entirely in the MSW region. However, the change from BP95 \cite{BP95} to
BP98 \cite{BP98} calculations increased the Homestake result to
about a third of the SSM
with a relatively large theoretical uncertainty.
Therefore it is quite possible that the ``MSW solutions'' extend to the dark
side as well.  Moreover, some people question the SSM
and/or the Homestake experiment, and perform fits by ignoring either
(or both) of them \cite{BHSSW}.  We show below that some of the MSW
solutions 
indeed extend to the dark side
and hence it is necessary to explore the dark side experimentally.  If
we further relax the theoretical prediction on the $^{8}$B solar
neutrino flux and/or ignore one of the solar neutrino experiments in
the global fit, the preferred regions extend even deeper into the dark
side.

Another possible reason for disregarding the dark side
is that the so-called ``vacuum oscillation
region'' ($\Delta m^{2} \lesssim 10^{-9}$~eV$^{2}$) was believed to be the
same in the light and dark sides.  This is because $P_{1}$ approaches
$\cos^{2} \theta$ for $\Delta m^{2} \ll 10^{-9}~{\rm eV}^{2} (E/{\rm
MeV})$ and Eq.~(\ref{eq:Pmatter}) reduces to Eq.~(\ref{eq:P}).  
It is remarkable, however, that low-energy
(especially \textsl{pp}) neutrinos do not reach this limit for $\Delta m^{2}
\gtrsim 10^{-10}~{\rm eV}^{2}$ and hence the preferred regions are
different in the light and the dark sides \cite{Alex}.
This observation also implies that the
separation of the MSW region 
and the vacuum oscillation region as traditionally
done in the global fits is artificial
and misleading.  It is important to study the entire range of
$\Delta m^{2}$ continuously.

If $\sin^2 2\theta$ is not a good parameter, what is the alternative?
Two suggestions have been made in the literature.  One is $\sin^2
\theta$, which is natural since the matter effect depends directly on
$\sin^2\theta$ \cite{us}.  If plotted on the linear scale, pure vacuum
oscillations would yield physics reflection-symmetric around $\sin^2
\theta=0.5$.  If plotted on the log scale, the reflection symmetry is
lost, but it is still a useful parameterization as physics is completely
continuous and smooth from the light to the dark side.  Another
possible parameterization is $\tan^2 \theta$, which retains the
reflection symmetry for pure vacuum oscillation around $\tan^2
\theta=1$ if plotted on the log scale \cite{FLMP,us}.  We employ
$\tan^2 \theta$ for the analysis below because we would like to use
the log scale to present the MSW solutions as well as the importance
of the matter effect on the ``vacuum oscillation'' region at the same
time.  Note that the Jacobian from $\sin^2 \theta$ or $\tan^{2}\theta$
to $\sin^2 2\theta$ is singular at $\theta = \frac{\pi}{4}$ and plots with
$\sin^2 2\theta$ will display unphysical singular behavior there
\cite{us}.

We next present the results of global fits to the current solar
neutrino data from water Cherenkov detectors (Kamiokande and
Super-Kamiokande) \cite{SKDN}, a chlorine target (Homestake) \cite{Cl}
and gallium targets (GALLEX and SAGE) \cite{Ga} on the full parameter
space. We do not
include the spectral data from Super-Kamiokande \cite{SKspec} as it
appears to be still evolving with time.  The fit is to the event rates
measured at these experiments only. In computing the rates we include
not only the \textsl{pp}, $^7$Be, and $^8$B neutrinos, but also the $^{13}$N,
$^{15}$O, and  \textsl{pep} neutrinos.
We use Eq.~(\ref{eq:Pmatter}) with $P_c$
computed in the exponential approximation for the electron number
density profile in the Sun, and properly account for neutrino
interactions in the Earth during the night with a realistic Earth
electron number density profile by numerically solving Schr\"odinger 
equation as described in \cite{us}.
Since the mixing angle at the production
point in the Sun's core depends on the electron number density, we integrate
over the production region numerically. 
We treat the
correlations between the theoretical uncertainties at different
experiments following Ref.~\cite{FLMP}.
To insure a smooth transition
between the MSW and the vacuum oscillation region, we 
integrate over the energy spectrum (including the thermal broadening 
of the $^{7}$Be neutrino ``line'')
for $\Delta m^{2} \leq 10^{-8}$~eV$^{2}$ and average the neutrino
fluxes over the seasons. For $\Delta m^{2} >10^{-8}$~eV$^{2}$ we treat
the two mass eigenstates as incoherent. Results are completely
smooth at $\Delta m^{2} = 10^{-8}$~eV$^{2}$, as expected. This allows
us to fit the data from $\Delta m^{2} = 10^{-11}$--$10^{-3}$~eV$^{2}$
all at once, unlike previous analyses which separate out the ``vacuum
oscillation region'' from the rest.

As was mentioned earlier, we take the global fit to the
currently available data only as indicative of the ultimate result
because we expect much better data to be collected in the near future to
eventually supersede the current data set.  We would like to
keep our minds open to surprises such as the possibility that one of the
earlier experiments was not entirely correct or that the theoretical
uncertainty in the flux prediction was underestimated.  In this
spirit, we employ more conservative attitudes in the global fit
than most of the analyses in the literature in the following three
possible ways.  (1) We allow higher confidence levels, such as
3~$\sigma$.  (2) We relax the theoretical prediction on the neutrino
flux.  (3) We ignore some of the experimental data in the fit.

\begin{figure}[tbp]
    \centerline{
    \psfig{file=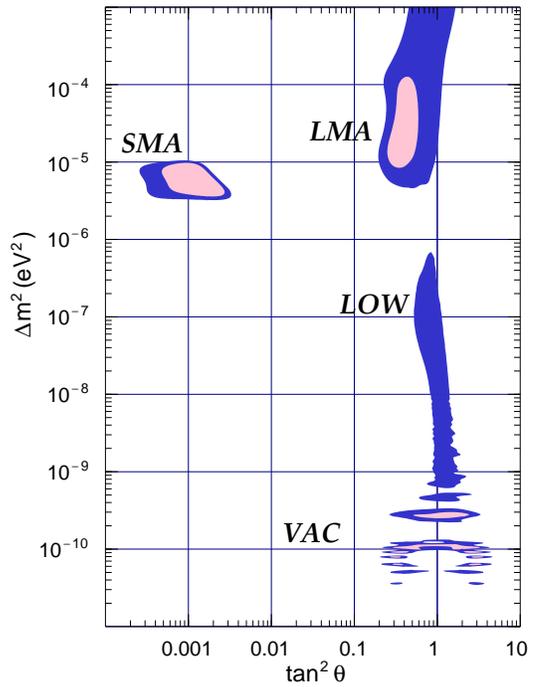,width=0.8\columnwidth}
        }
    \caption{A global fit to the solar neutrino event rates at
      chlorine, gallium and water Cherenkov experiments.  The regions
      are shown at 2~$\sigma$ (light shade) and 3~$\sigma$ (dark
      shade) levels.  The region $\tan^{2} \theta > 1$ is the dark
      side $\theta > \frac{\pi}{4}$.}
    \label{fig:standardfit}
\end{figure}

The global fit results are presented in Fig.~\ref{fig:standardfit} at
the 2~$\sigma$ (95\% CL) and 3~$\sigma$ (99.7\% CL) levels defined by
$\chi^{2} - \chi^{2}_{\it min}$ for two degrees of freedom.  It is
noteworthy that both the LMA and LOW solutions (we use the
nomenclature introduced in
\cite{BKS}) extend to the dark side at the 3~$\sigma$ level.  At 99\%
CL, however, the LMA solution is confined to the light side.  This
result is consistent with the two-flavor limit of the three-flavor
analysis in \cite{FLMP} and the four-flavor analysis in
\cite{4flavor}, where the spectral data is included and the LOW
solution extends into the dark side at 99\% CL.  Another interesting
fact is that the LOW solution is smoothly connected to the VAC
solution, where the preferred region is clearly asymmetric between the
light and the dark sides.  Note that, at $\Delta m^{2} \sim 
10^{-9}~{\rm eV}^{2}$, the allowed
region is bigger in the dark side.
The region $10^{-9} < \Delta m^{2} < 10^{-8}~{\rm eV}^{2}$
was, to the best of the authors' knowledge, never studied
fully in the literature and this result demonstrates the need to study
the entire $\Delta m^{2}$ region continuously without the artificial
separation of the ``MSW region'' and ``vacuum oscillation region,'' as
traditionally done in the literature.

\begin{figure}[tbp]
    \centerline{
    \psfig{file=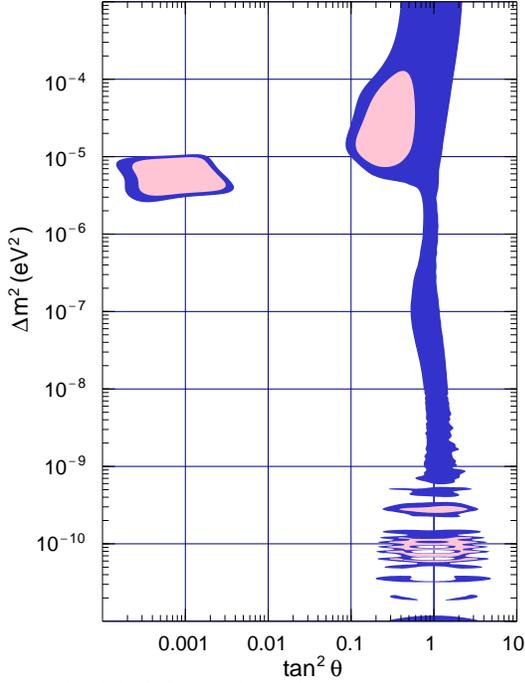,width=0.8\columnwidth}
        }
    \caption{A global fit to the solar neutrino event rates at
      chlorine, gallium and water Cherenkov experiments, where the
      $^{8}$B flux is treated as a free parameter.  Contours are shown
      at 2~$\sigma$ (light shade) and 3~$\sigma$ (dark shade).}
    \label{fig:freeB8}
\end{figure}

We next present a fit where the theoretical prediction of the $^{8}$B
flux is relaxed.  Even though the helioseismology data constraints the
sound speed down to about 5\% of the solar radius \cite{BP98}, the
core region where $^{8}$B neutrinos are produced is still not
constrained directly.  Given the sensitive dependence of the $^{8}$B
flux calculation on the core temperature $\Phi_{{}^{8}B} \sim T^{22}$, we
may consider it as a free parameter in the fit.  This can be done
within the formalism of Ref.~\cite{FLMP} by formally sending the error
in $C_{\rm Be}$ to infinity.  The result is presented in
Fig.~\ref{fig:freeB8}.  The preferred region
extends more into the dark side than the previous fit.  
Even though the LMA and
LOW solutions are connected in this plot, the lack of a large day-night
asymmetry at Super-Kamiokande would eliminate the range $3 \times
10^{-7} \lesssim \Delta m^{2} \lesssim 10^{-5}~{\rm eV}^{2}$ for
$0.2 \lesssim \tan^{2} \theta < 1$ \cite{SKDN}.  It is important for
Super-Kamiokande to report their exclusion region on the dark side.

\begin{figure}[tbp]
    \centerline{
    \psfig{file=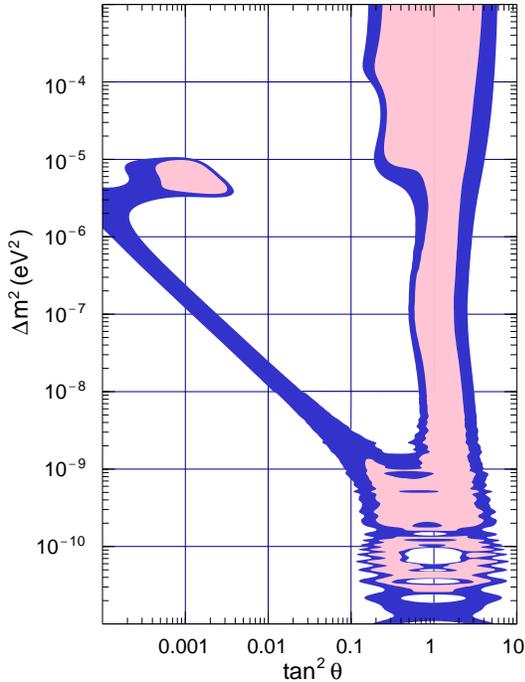,width=0.8\columnwidth}
        }
    \caption{A global fit to the solar neutrino event rates at the gallium
      and water Cherenkov experiments but not at the chlorine
      experiment.  Contours are shown at 2~$\sigma$ (light shade) and
      3~$\sigma$ (dark shade).}
    \label{fig:nochlorine}
\end{figure}

Finally, we present a fit where the event rate measured at the Homestake
experiment is not used in Fig.~\ref{fig:nochlorine}.  This may be a
sensible exercise given that the neutrino capture efficiency was
never calibrated in this experiment.  The preferred region extends
into the dark side even at the 95\% CL.  Note also the asymmetry between
the dark and the light sides even for $\Delta m^{2} < 10^{-9}~{\rm eV}^{2}$.

We expect the data of the current and next generation of solar neutrino
experiments, such as Super-Kamiokande, SNO, GNO, Borexino, KamLAND, to
eventually supersede the current data set.  Therefore we regard the
above global fits only as estimates of the ultimate results.  The
most important point is that all experimental collaborations should
report their results, both exclusion and measurements, on both sides
of the parameter space, without unnecessary theoretical bias towards
the light side.  We strongly urge the experimental collaborations to
consider this point.


\begin{thebibliography}{99}
\bibitem{decoherence} For instance, integration
over neutrino energies washes out the oscillatory factor.
In the case of $^{7}$Be neutrinos, this occurs
for $\Delta m^{2} \gtrsim 10^{-8}$~eV$^{2}$ 
due to the thermal broadening effect.  For neutrinos with
continuous spectra, the washout occurs at even lower $\Delta
m^{2}$.  See \cite{seasonal,Alex} for more details.
\end{thebibliography}
\end{document}